\documentclass[3p,times,10pt,number,sort&compress]{elsarticle}
\usepackage{mathptmx} % times font

\usepackage[T1]{fontenc}
\usepackage[latin9]{inputenc}
\usepackage{textcomp}
\usepackage{microtype}
\usepackage{color}
\usepackage{amsthm}
\usepackage{amssymb}

\usepackage{graphicx}
\usepackage{epstopdf}
\newcommand{\mycomment}[1]{}

\theoremstyle{plain}

\newcommand{\drafttitle}{}
\newcommand{\draftpageheader}{}

\newcommand{\commentfig}[1]{#1}
\def \plotheight {4cm}

\journal{Commun Nonlinear Sci Numer Simulat}

\begin{document}

\begin{frontmatter}

\title{\drafttitle 
Quasi-separatrix layers and three-dimensional reconnection diagnostics for line-tied tearing modes}

\author{A.~S.~Richardson}
\ead{asrichardson@lanl.gov}
\author{J.~M.~Finn}
\ead{finn@lanl.gov}

\address{T-5, Applied Mathematics and Plasma Physics, Los Alamos National Laboratory,
Los Alamos, NM, USA}

\begin{abstract}
In three-dimensional magnetic configurations for a plasma in which no closed field line
or magnetic null exists, no magnetic reconnection can occur, by the
strictest definition of reconnection. A finitely long pinch with line-tied
boundary conditions, in which all the magnetic field lines start at
one end of the system and proceed to the opposite end, is an example
of such a system. Nevertheless, for a long system of this type, the
physical behavior in resistive magnetohydrodynamics (MHD) essentially involves reconnection. This has been
explained in terms comparing the geometric and tearing widths \cite{delzanno:032904,huang:042102}. The concept of a quasi-separatrix
layer\cite{Priest:1995fk,1996A&A...308..643D} was developed for such systems.
In this paper we study a model for a line-tied system in which the
corresponding periodic system has an unstable tearing mode. We analyze
this system in terms of two magnetic field line diagnostics, the \emph{squashing
factor}\cite{Priest:1995fk,1996A&A...308..643D,Titov20021087} and the  electrostatic potential difference used
in kinematic reconnection studies\cite{1990ApJ...350..672L,1991ApJ...366..577L}. We discuss
the physical and geometric significance of these two diagnostics and
compare them in the context of discerning tearing-like behavior in line-tied modes.
\end{abstract}

\begin{keyword}
Magnetic reconnection \sep tearing \sep line-tying \sep quasi-separatrix layers

\end{keyword}

\end{frontmatter}

\draftpageheader

\section{Introduction}

It has been argued that in three dimensions hyperbolic closed field lines (X-lines)\cite{1990ApJ...350..672L}
or magnetic nulls\cite{1988JGR....93.8583G,1990ApJ...350..672L} are required for magnetic reconnection.  In two dimensions, reconnection has also typically been discussed in terms of nulls (zero guide field) or X-lines (nonzero guide field).  See Ref.~\cite{NR:BirnPriestBook} and references therein.
In this strictest definition of reconnection, the argument is this:
in ideal magnetohydrodynamics (MHD) these structures can lead to singularities
which are then resolved by non-ideal effects, and in the absence of
such structures, no singularities occur. In the presence of small electrical resistivity $\eta$,
these singularities become current sheets of finite but small thickness dependent on $\eta$.
For our purposes here, we
will describe the non-ideal effects as arising from finite electrical
resistivity $\eta$, i.e.~we consider reconnection in resistive MHD.\footnote{However, there is no fundamental reason why these concepts cannot
apply to reconnection with other non-ideal effects, such as electron
inertia and hyper-resistivity.}  In this
strict definition, the question of whether reconnection occurs is
in the context of the limit $\eta\rightarrow0$, or the Lundquist
number $S\rightarrow\infty$. 

In a line-tied system of finite length $-L\leq z\leq L$ with 
fields having $B_{z}>0$, all field lines enter at
$z=-L$ and exit at $z=L$. Therefore, closed field lines cannot form. (We will
speak only in terms of closed field lines; in order for nulls
to form in this geometry, all components of ${\bf B}$ would have to change sign in the region.)
So in the strict
sense described above, no magnetic reconnection can occur. However,
in Refs.~\cite{delzanno:032904,huang:042102} line-tied modes which are tearing modes
in cylindrical geometry [for equilibria with $B_{\theta 0}(r)$, $B_{z0}(r)$] with $L=\infty$
were studied. In these $L=\infty$ modes, behaving as $e^{im\theta+ikz}$,
there is a mode rational surface consisting of closed field lines
in the equilibrium, and the perturbed fields have hyperbolic
and elliptic closed field lines (islands). The conclusion of these papers
was that the modes still behave as tearing modes (with growth rates
$\gamma$ proportional to a fractional power of $\eta$) in a sufficiently
long line-tied system $L>L_{\rm crit}$. The condition $L>L_{\rm crit}$
is defined as the range over which the tearing width $w_{t}$ of the
mode is greater than the geometric width $w_{g}$, the width of the
mode associated with the line-tying boundary condition [$w_g\propto 1/L$]. For $L<L_{\rm crit}$,
when $w_{t}<w_{g}$, the modes no longer resemble tearing modes; they have
no recognizable tearing layer and $\gamma\propto\eta$, i.e.~they involve
global resistive diffusion rather than reconnection.

These results suggest that in many cases it may make more sense to
define reconnection for fixed but large $S$, rather than in the limit
$S\rightarrow\infty$. This is particularly true for numerical simulations with limited values of the Lundquist number $S$ and for laboratory experiments with relatively low temperatures. Three such experiments suited for reconnection studies with line-tying are RWM\cite{PhysRevLett.96.015004}, RSX\cite{furno:2324}, and LAPD\cite{4989217}.  This point of view is consistent with that expressed in
Ref.~\cite{Priest:1995fk,1996A&A...308..643D}, in which the concept of a \emph{quasi-separatrix
layer} (QSL) was introduced for fields without hyperbolic closed field
lines (or nulls) as a resolution of controversies associated with generalized magnetic reconnection\footnote{The controversies involving GMR related to whether localized current had to be from reconnection or could be due to double layers, Pfirsch-Schluter currents, or other sources.}\cite{Hesse:1988rt,Schindler:1988yq}. In fact, the concept of a QSL was used to analyze LAPD in \cite{PhysRevLett.103.105002}. For such magnetic fields (with large but fixed $S$), the geometric aspects of the field lines can act to separate the field lines in a manner qualitatively similar to fields with hyperbolic lines. If this separation occurs in a thin enough region, the physics is basically identical to the behavior in the presence of a hyperbolic line.
However, note that while the condition $w_g <w_t$ may be satisfied for such laboratory systems and for simulations with relatively small $S$, for solar coronal or astrophysical applications this requirement may hold only for unrealistically long plasmas.

In Refs.~\cite{Priest:1995fk,1996A&A...308..643D,ESASpec.Publ.448:7151999,Titov20021087,Titov:2002fj,0004-637X-660-1-863,0004-637X-693-1-1029}, it was suggested that the most effective measure of a QSL is the \emph{squashing factor}, which measures the squeezing and stretching of field lines. 
The method of slip-squashing factors\cite{0004-637X-693-1-1029} also takes resistivity (or other non-ideal effects) into account. It involves computing the squashing factor plus taking into account the field line slippage due to non-ideal effects. For fields which are determined by boundary motions, this slippage is measured by comparing the initial field line mapping with the field line mapping at a later time.

In Sec.~\ref{sec:diagnostics} of this paper we discuss the squashing factor $Q$ and
a second diagnostic, related to the potential $\phi_i$ used in Refs.~\cite{1990ApJ...350..672L,1991ApJ...366..577L}.
The latter function is the scalar potential required to give $\mathbf{E}\cdot\mathbf{B}=0$,
a consequence of Ohm's law in ideal MHD. We first compute $Q$ and
$\phi_i$ for two examples of magnetic fields, namely a doublet-like
field and a field given by an equilibrium $B_{\theta0}(r)$, $B_{z0}(r)$
plus a single tearing mode, both in a finite region $-L<z<L$.  We also compute $\phi_r$, the scalar potential required to satisfy ${\bf E}\cdot \mathbf{B}=\eta {\bf j}\cdot \mathbf{B}$ in resistive MHD. In Sec.~3 we review the treatment of
line tied modes in the \emph{two-mode approximation}\cite{evstatiev:072902}, which applies for long plasmas. In Sec.~4 we study QSLs in
a model representing growing tearing modes with line-tied boundary
conditions in the two-mode approximation. We show the squashing factor
$Q$ as well as the potentials $\phi_i$ and $\phi_r$ for modes of various amplitudes, in the two cases $w_t>w_g$ and $w_t < w_g$.
In Sec.~5 we summarize and discuss our results. In the Appendix we
discuss other mathematical issues related to the squashing factor.

\section{Reconnection diagnostics}\label{sec:diagnostics}

In this section we describe two diagnostics that can be used to identify reconnection and QSLs in systems with finite resistivity.  The first diagnostic is the {\em squashing factor}, defined in Refs.~\cite{ESASpec.Publ.448:7151999,Titov20021087,Titov:2002fj,0004-637X-660-1-863}, and the second is the potential difference $\Delta\phi$ which is required to satisfy either the ideal MHD relation ${\bf E}\cdot{\bf B}=0$ (see Ref.~\cite{1990ApJ...350..672L}) or the resistive relation ${\bf E}\cdot{\bf B}=\eta {\bf j}\cdot {\bf B}$.  Each of these diagnostics require computing the magnetic fields lines which run the length of the system, from $z=-L$ to $z=L$.

\subsection{Squashing Factor}
The first diagnostic we consider is the squashing factor.
Integrating field lines from one end of the system to the other defines a coordinate mapping $M:{\bf x}\to {\bf X}$, where $\bf x$ is the starting point of the line in the $z=-L$ plane, and $\bf X$ is the ending point in the $z=L$ plane.  (In the more general case, we map from the set with normal field $B_n<0$ to the set with $B_n>0$.) Geometrically, stretching and squashing of flux tubes indicates a potential for reconnection, and the Jacobian $J$ of the mapping $M$ gives us information about these processes. The overall expansion of flux tubes is not associated with reconnection, but rather is related to variation of the guide field strength $B_z$. Specifically, flux conservation implies $B_z (x_1,x_2,-L)dx_1 dx_2=B_z (X_1,X_2,L)dX_1 dX_2$, so that the Jacobian determinant $D\equiv \det(J)$ satisfies $D=B_z (x_1,x_2,-L)/B_z (X_1,X_2,L)$.  The mapping $M$ takes a flux tube with a circular cross-section and ``squashes'' it so that its cross-section is elliptical. In order to measure the stretching and squashing while compensating for the expansion of flux tubes if $B_z \neq {\rm const.}$, a natural quantity to consider\cite{Priest:1995fk,1996A&A...308..643D,ESASpec.Publ.448:7151999,Titov20021087,Titov:2002fj,0004-637X-660-1-863,0004-637X-693-1-1029} is the aspect ratio of the elliptical cross section of the flux tube at $z=L$.\footnote{In the Appendix, we discuss the possibility of transforming $(x_1,x_2)$ and $(X_1,X_2)$ to canonical variables, for which $D=1$. }
\begin{figure}[htbp]
\begin{center}
\commentfig{
\includegraphics[width=5cm]{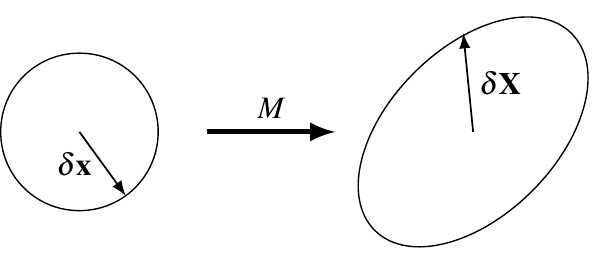}
}
\caption{Squashing of flux tube cross-section by the mapping $M$. The major and minor axis of the ellipse have lengths $\rho_{max}=\max(|\delta{\bf X}|)$ and $\rho_{min}=\min(|\delta{\bf X}|)$ (where $\rho_i$ are the singular values of the matrix $J$), and so the aspect ratio of the ellipse is $R={\rho_{max}}/{\rho_{min}}$. This ratio removes the effect of variations in $B_z$, which would only give an overall expansion or contraction of the flux tube.\label{fig:squash}}
\end{center}
\end{figure}

Consider two field lines whose initial points are separated by the tangent vector $\delta{\bf x}$.  Their endpoints will be separated by $\delta{\bf X}$, and the length of $\delta{\bf X}$ compared to the length of $\delta{\bf x}$ will tell us about the stretching of a flux tube containing the two field lines.  The lengths of these two vectors can be directly compared since the Jacobian matrix $J$ ($J_{ij}=\partial X_i / \partial x_j$) maps tangent vectors, i.e.~$\delta\mathbf{X}=J\delta\mathbf{x}$. Thus, their lengths are related by
\begin{eqnarray}\label{eq:lengths}
(\delta\mathbf{X},\delta\mathbf{X})=(J\delta\mathbf{x},J\delta\mathbf{x})=(\delta\mathbf{x},J^{T}J\delta\mathbf{x}).
\end{eqnarray}
The symmetric positive definite matrix $J^{T}J$ is the covariant metric tensor on the surface at
$z=-L$ derived from a Euclidean metric tensor on the surface at $z=L$, and its eigenvalues $\lambda_\pm$, obtained from the Rayleigh quotient $(\delta\mathbf{x},J^{T}J\delta\mathbf{x}) /(\delta\mathbf{x},\delta\mathbf{x})$, tell us how the shape of a flux tube changes.\footnote{The quantity ${\rm Tr}(J^TJ)$ was described as the norm of the displacement gradient tensor in \cite{Priest:1995fk}. It can also be written as $\vert \nabla X_1 \vert^2 + \vert \nabla X_2 \vert^2$ and is actually the square of the Frobenius norm of $J$.}
In fact, the ratio of the singular values\cite{NR:svd_evals} $\rho_\pm = \sqrt{\lambda_\pm}$ equals the aspect ratio of the ellipse.  See Fig.~\ref{fig:squash}. 
This ratio is easily shown to obey
\begin{eqnarray}\label{eq:squashing}
R=\frac{\rho_{max}}{\rho_{min}} = \frac{Q}{2} + \sqrt{ \left(\frac{Q}{2}\right)^2 - 1},\quad{\rm  where}\quad Q \equiv T/D = {\rm Tr}(J^T J) / |\det(J)|.
\end{eqnarray}
This quantity $Q$, which also equals $R+1/R$, is called the {\em asymptotic squashing factor}\cite{0004-637X-660-1-863}, or just {\em squashing factor}.
In regions where $Q$ is large, $R=Q-1/Q+\cdots \approx Q$. It was argued in Refs.~\cite{Priest:1995fk,1996A&A...308..643D,ESASpec.Publ.448:7151999,Titov20021087,Titov:2002fj,0004-637X-660-1-863,0004-637X-693-1-1029} that these regions, where the stretching and squashing are large, are candidates for the occurrence of reconnection.   Such regions of large $Q$ are called {\em quasi-separatrix layers}. We are interested in only these regions and, following Refs.~\cite{Priest:1995fk,1996A&A...308..643D,ESASpec.Publ.448:7151999,Titov20021087,Titov:2002fj,0004-637X-660-1-863,0004-637X-693-1-1029}, we will thus focus on $Q$ rather than $R$. In the next section, we give an example which illustrates how $Q$ becomes larger and more concentrated for larger $L$.

\subsection{Squashing Factor Example: Doublet Field}\label{sec:Q_example}

\begin{figure}[htbp]
\begin{center}
\commentfig{
\includegraphics[]{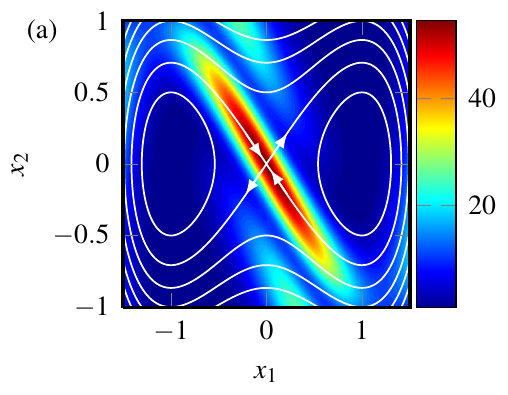}
\includegraphics[]{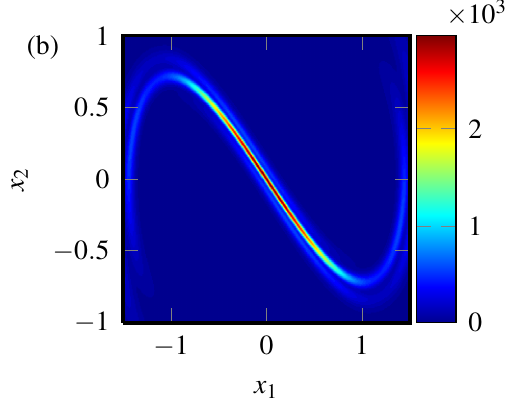}
\includegraphics[]{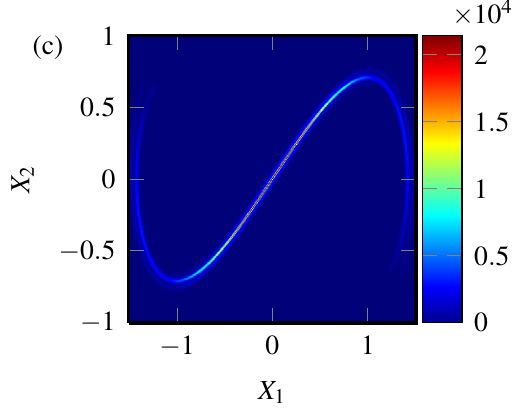}
%\tikzset{external/remake next}
%\input{Q_db_2.pgf}
%\tikzset{external/remake next}
%\input{Q_db_4.pgf}
%\tikzset{external/remake next}
%\input{Q_db_5_end.pgf}
}
\caption{Squashing factor $Q(x_1,x_2)$ for the model of Eq.~(\ref{eq:doublet}), with $B_0=1$ and (a)~$L=1$,  (b)~$L=2$.  In (c), the squashing factor for $L=2.5$ is shown plotted as a function of ending point $\bf X$.
In (a), contours of $A_z$ are shown, with arrows indicating the stable and unstable manifolds of the X-line at the origin.  In this 2D case, the stable and unstable manifolds coincide to form a separatrix.
The quasi-separatrix layer traces out the {\em stable} manifold when plotted as a function of initial point $\bf x$ [(a) and (b)] and the {\em unstable} manifold when plotted as a function of the ending point $\bf X$ in (c).
\label{fig:db_Q}}
\end{center}
\end{figure}

Consider the vector potential with $A_{x_1} = -B_0 x_2$, $A_{x_2} = 0$, and $A_z = \frac{x_2^2}{2} -\frac{x_1^2}{2} + \frac{x_1^4}{4}$.  This defines the magnetic field
\begin{eqnarray}\label{eq:doublet}
B_{x_1} = {x_2}, \quad 
B_{x_2} = {x_1} - x_1^3, \quad
B_z = B_0.
\end{eqnarray}
The equations for field lines are then
\begin{eqnarray}
{d{x_1}}/{dz} = {x_2}/B_0, \quad
{d{x_2}}/{dz} = ({x_1}-x_1^3)/B_0.
\end{eqnarray}
These equations are in canonical form, with Hamiltonian $H=A_z(x_1,x_2)/B_0$ and $D=1$  (since $B_z=B_0={\rm const.}$). The contours of $A_z$ for this field are shown in Fig.~\ref{fig:db_Q}a. Integrating the field lines from $z=-L$ to $z=L$ with a symplectic integrator
determines an exactly area preserving  mapping $(x_1,x_2,-L) \to (X_1,X_2,L)$.  The Jacobian matrix of this mapping is estimated numerically by tracing field lines for a grid of initial points ${\bf x}$, and taking the difference of the ending points ${\bf X}$. Since this model is two dimensional, we can change the length $L$ without modifying the model, as we do in the 3D models in Sec.~\ref{sec:line-tied}.  The squashing factor $Q$ as a function of $(x_1,x_2)$ is shown in Fig.~\ref{fig:db_Q} for lengths $L=1$, $2$, and $2.5$. Note that for $L=2$, $Q$ is peaked in an elliptically shaped region near the part of the stable manifold near the X-line. As $L$ increases, $Q$ becomes much more peaked and concentrated near the stable manifold and further along the stable manifold. The squashing factor $Q$ is shown in Fig.~\ref{fig:db_Q}c for $L=2.5$ as a function of the variables at $z=L$, namely $(X_1,X_2)$. The large values of $Q$ in this figure are concentrated on the {\em unstable} manifold because this same quantity $Q(X_1,X_2)$ could also have been computed by initializing at $(X_1,X_2,L)$ at $z=L$ and integrating backwards to $(x_1,x_2,-L)$.

\begin{figure}[htbp]
\begin{center}
\commentfig{
%\myfig{Q2_3D.pdf}
\includegraphics[height=4.7cm]{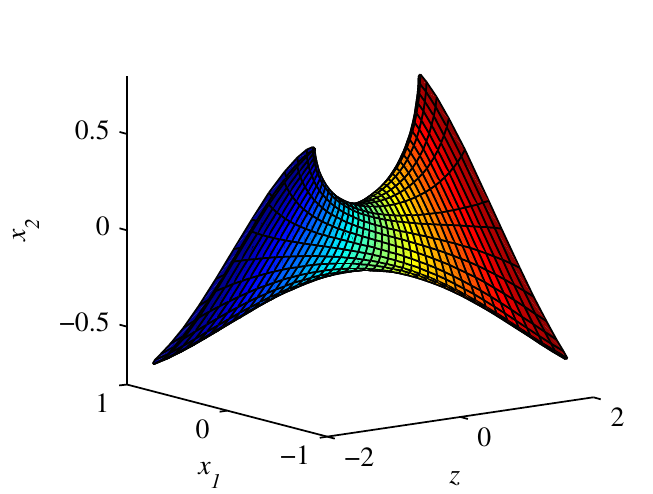}
}
\caption{3-dimensional visualization of the quasi-separatrix layer for the model of  Eq.~(\ref{eq:doublet}).  The surface is formed by the field lines which have $Q= 1000$ for $L=2$.
\label{fig:db_Q_3D}}
\end{center}
\end{figure}

In Fig.~\ref{fig:db_Q_3D} we show a three dimensional plot of the contour where $Q=1000$ for $L=2$. The concentration along the stable manifold for $z=-L$ and along the unstable manifold for $z=L$ is evident.  This surface has the form of a hyperbolic flux tube, which arises in models of reconnection in solar physics, and the study of QSLs.\cite{Titov:2002fj,APD:currentsandHFT,0004-637X-582-2-1172}

\mycomment{%%%
\begin{figure}[htbp]
\begin{center}
\commentfig{
\includegraphics[height=\plotheight]{Q5.pdf}
\includegraphics[height=\plotheight]{Q5_rev.pdf}
}
\caption{Squashing factor $Q$ for for the model of Eq.~(\ref{eq:doublet}), with $B_0=1$ and $L=2.5$, plotted as a function of (a) the initial point $\bf x$ of the field line and (b) the ending point $\bf X$ of the field line.  The quasi-separatrix layer traces out the {\em stable} manifold and the {\em unstable} manifold of the X-line at the origin in (a) and (b), respectively.}
\label{fig:db_Q_rev}
\end{center}
\end{figure}
}%%%

\subsection{Scalar Potential Difference $\Delta \phi$}

The second diagnostic we consider in this paper is the potential difference between the two ends of the field lines. We calculate $\Delta \phi=\Delta \phi_i$ from the ideal MHD Ohm's law $\mathbf{E}+\mathbf{v}\times\mathbf{B}=0$ to give zero
parallel electric field $ E_{||}\propto\mathbf{E}\cdot\mathbf{B}$, and calculate $\Delta \phi_r$ similarly from the resistive MHD Ohm's law $\mathbf{E}+\mathbf{v}\times\mathbf{B}=\eta {\bf j}$. (Again, we consider resistive MHD just for concreteness.) 
The first approach, for ideal MHD, was used to analyze three dimensional reconnection in
the presence of hyperbolic closed field lines or nulls in Ref.~\cite{1990ApJ...350..672L}.
In this paper, we compute both $\Delta \phi_i$ and $\Delta \phi_r$ by taking the magnetic field $\mathbf{B}(\mathbf{x},t)$ and the inductive electric field $-\partial\mathbf{A}/\partial t$ computed from resistive MHD. We solve
$\mathbf{E}\cdot\mathbf{B}=0$ and $\mathbf{E}\cdot\mathbf{B}=\eta {\bf j}\cdot {\bf B}$ along field lines, namely 
\begin{equation}
\mathbf{B}\cdot\nabla\phi_i=-\mathbf{B}\cdot\frac{\partial\mathbf{A}}{\partial t};\,\,\,\,\,\,\,\,\,\, \mathbf{B}\cdot\nabla\phi_r=-\mathbf{B}\cdot\frac{\partial\mathbf{A}}{\partial t} -\eta {\bf j}\cdot {\bf B}, \label{eq:BdotGradPhi}
\end{equation}
for the scalar potentials $\phi_i$, $\phi_r$ and, for both, evaluate the difference between the potential at the two ends of the field line at $z=-L$ and $z=L$.  
The potential $\phi_r$ is exactly the single-valued scalar potential related to the fields obtained in resistive MHD (in a specific gauge) and represents the electrostatic potential due to small (quasineutral) charge variations required to balance the inductive electric fields of the mode locally along the field lines, in the presence of the resistive term $\eta {\bf j}\cdot {\bf B}$.

In computing $\phi_i$, we analyze the fields generated by resistive MHD in the context of \emph{ideal} MHD, in order to determine when $\Delta \phi_r \approx \Delta \phi_i$. In most regions of the plasma, the potential $\phi_i$ is also a single-valued function, again due to movement of electrons along ${\bf B}$ to cancel the inductive electric field. In regions where $\eta {\bf j}\cdot {\bf B}$ is small, $\Delta \phi_r \approx \Delta \phi_i$. On the other hand, if there is a closed field line in the system
and if $\partial\mathbf{A}/\partial t$ integrates to a nonzero value along this
line, then mulit-valuedness of $\phi_i$ will occur at points on the line and singular behavior will occur on
points that go to the line (the {\em stable manifold} of the hyperbolic X-line), because the associated flux change through any
surface bounded by the line cannot be cancelled by a single-valued electrostatic potential.\footnote{Note that a gauge change $\mathbf{A}\rightarrow\mathbf{A}+\nabla\chi,\,\,\phi\rightarrow\phi-\partial\chi/\partial t$
with $\chi$ continuous does not affect the singular nature of $\phi_i$
as found from Eq.~(\ref{eq:BdotGradPhi}). Similarly, in integrating
Eq.~(\ref{eq:BdotGradPhi}) from $z=-L$ to $z=L$, a smooth initial value $\phi(r,\theta,z=-L)$ is irrelevant.}
 In the line-tying geometry of this paper, we do not expect to encounter true singularities
in $\phi_i$. However, we do find that $\phi_i$, like the squashing
factor $Q$, can be large and strongly localized, which is indicative
of reconnection-like behavior, e.g.~current sheets. This kind of behavior, with peaked but nonsingular values of $\phi_i$, was seen and discussed in \cite{1991ApJ...366..577L}.

It is useful to compute both $\phi_i$ and $\phi_r$, since the difference $\Delta\phi_i - \Delta\phi_r$ is simply the integral of $\eta {\bf j}\cdot {\bf B} \propto \eta j_{||}=E_{||}$ along the field line.  Parallel currents are associated with reconnection effects, and this method allows us to not only compute the parallel currents (through $\Delta\phi_i - \Delta\phi_r$), but it also gives us quantities ($\Delta \phi_i$ and $\Delta \phi_r$) with which to compare the parallel currents.  If $\Delta \phi_r\approx\Delta \phi_i$, then the effect of the parallel currents are small compared to the inductive electric fields.  In regions where $\Delta\phi_i - \Delta\phi_r$ is large compared to $\Delta\phi_i$, however, the currents --  and thus reconnection effects -- are important. 
The difference $\Delta\phi_i -\Delta\phi_r$ is equal to the quasi potential $\Xi$ of Ref.~\cite{0004-637X-631-2-1227}. However, we find that comparing the two separate potentials $\Delta\phi_i$ and $\Delta\phi_r$ is more informative than considering only $\Xi = \Delta\phi_i -\Delta\phi_r$.

As diagnostics for reconnection, there is one fundamental difference between the 
squashing factor $Q$ and the potentials $\phi_i$, $\phi_r$: The former involves
only the structure of the magnetic field at a specific time, whereas
the potentials $\phi_i$, $\phi_r$ obtained from Eq.~(\ref{eq:BdotGradPhi}) involve
the structure of the magnetic field \emph{and} the inductive electric
field $-\partial\mathbf{A}/\partial t$. That is, an area of large
and localized $Q$ only indicates the possibility of having reconnection,
whereas a large and localized value of $\Delta\phi_i$ shows flux changes
{\em and} field line topology indicative of reconnection.

\subsection{Ideal MHD Scalar Potential Example: Single tearing Mode}

We use the compressible zero-$\beta$ visco-resistive MHD equations:
\begin{eqnarray}
\partial_t \rho  &=& - \nabla\cdot ( \rho{\bf v})\\
\rho \partial_t {\bf v} +\rho {\bf v} \cdot \nabla {\bf v} &=& {\bf j} \times {\bf B} + \nu\rho\nabla^2{\bf v}\\
{\bf E} + {\bf v}\times {\bf B} &=& \eta {\bf j} \label{eq:ohm}\\
\nabla\times{\bf E}&=&-\partial_t {\bf B}\\
\nabla\times{\bf B} &=& {\bf j}\\
\nabla\cdot{\bf B} &=& 0,
\end{eqnarray}
in a cylinder with $-L\leq z\leq L$ and $0\leq r\leq r_w$, with $r_w=2$.

The equilibrium used was the equilibrium of \cite{delzanno:032904}, for which the fastest growing mode in an infinite cylinder is a tearing mode rather than an ideal MHD mode. This force-free equilibrium is specified by its axial current density
$
j_{z0}(r)={2}/({1+r^{\kappa}})
$
with $\kappa=6$, and $B_{z0}(r)$ determined by force balance with an integration constant $B_{z0}(r=0)=5$, with $\rho_0 =1$ and ${\bf v}_0 = 0$. This equilibrium is ``tokamak-like'' in the sense of having an increasing profile of the field line pitch $\mu(r)\equiv rB_{z0}(r)/B_{\theta 0}(r)$ and $B_{z0} \gg B_{\theta 0}$.

The linearized MHD equations are
\begin{eqnarray}
\partial_t \tilde{\bf v} &=& \left[ \nabla \times \tilde{\bf B} - \lambda(r) \tilde{\bf B}  \right] \times {\bf B}_0 + \nu \nabla^2 \tilde{\bf v} \label{eq:mhdlin1}\\
\partial_t \tilde{\bf B} &=& \nabla \times \left( \tilde{\bf v} \times {\bf B}_0 - \eta \nabla \times \tilde{\bf B} \right),\label{eq:mhdlin2}
\end{eqnarray}
with $\tilde{{\bf B}}=\tilde{{\bf B}}(r)e^{im\theta +ikz}$ and $\tilde{{\bf v}}=\tilde{{\bf v}}(r)e^{im\theta +ikz}$.  These equations are solved\footnote{As in Ref.~\cite{richardson:112511}, a small amount of divergence cleaning is used.} as an eigenvalue problem for the growth rate $\gamma$, with $m=1$, and $k=-0.13507$, giving a mode rational surface (where ${\bf k}\cdot{\bf B}_0=0$) at $r_s=1.2576$.  For computing the field line diagnostics, we have superimposed the equilibrium fields with the perturbed fields multiplied by a small amplitude $a$.\footnote{For even relatively small values of the amplitude $a$, nonlinear effects could be important, invalidating the assumption that we can superimpose the equilibrium and perturbed fields obtained by linear theory. For our purposes in this paper, we will keep $a$ small so that such errors are small.}  The system length $L$ was taken to be $L=20\pi/k$.  This length and the value of $k$ were chosen for comparison with the results from the two-mode approximation in the next section.

\begin{figure}[htbp]
\begin{center}
\commentfig{
\includegraphics[]{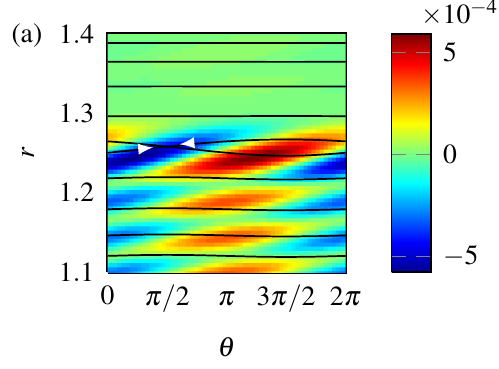}
\includegraphics[]{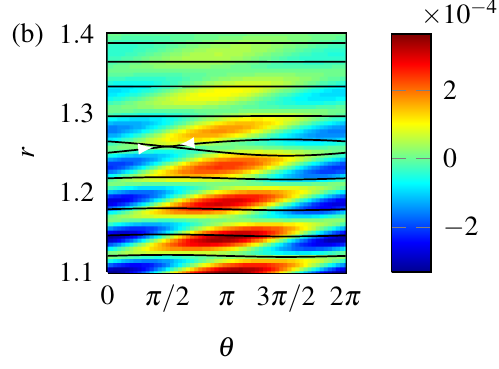}
\includegraphics[]{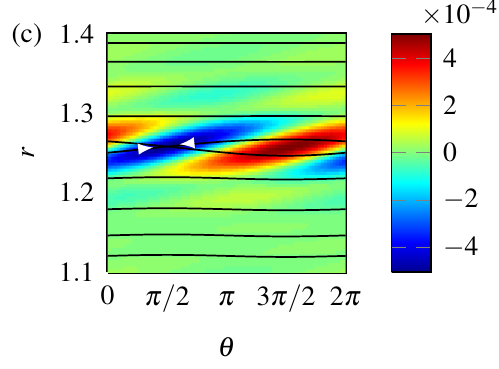}
%\tikzset{external/remake next}
%\input{newfigs3/phi_i.pgf}
%\tikzset{external/remake next}
%\input{newfigs3/phi_r.pgf}
%\tikzset{external/remake next}
%\input{newfigs3/phi_diff.pgf}
}
\caption{(a) $\Delta \phi_i(r,\theta)$, and  (b) $\Delta \phi_r(r,\theta)$, computed by integrating Eq.~(\ref{eq:BdotGradPhi}) from $z=-L $ to $z=L$.  Mode amplitude $a=1\times 10^{-3}$. (c) Difference between $\Delta \phi_i$ and $\Delta \phi_r$, which is the cumulative parallel current integrated along the field lines. Contours are of the helical flux, which for one mode is identical to the surface of section. The white arrows indicate the direction of the stable manifold of the X-line near $(r,\theta)=(1.25,\pi/2)$. 
\label{fig:1mode_phi}}
\label{default}
\end{center}
\end{figure}

\begin{figure}[htbp]
\begin{center}
\commentfig{
\includegraphics[]{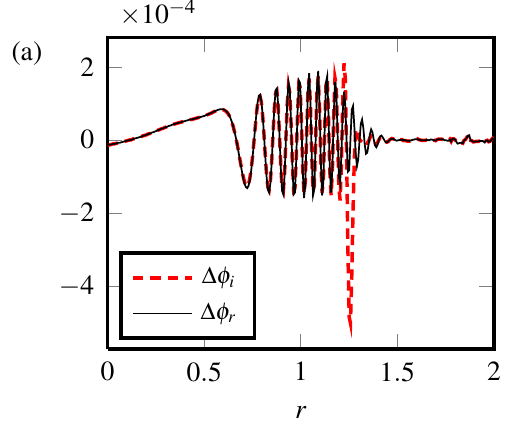}
\includegraphics[]{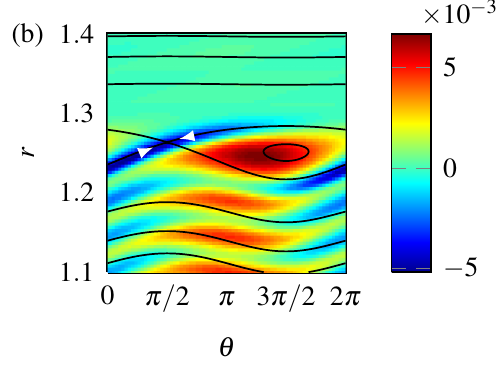}
\includegraphics[]{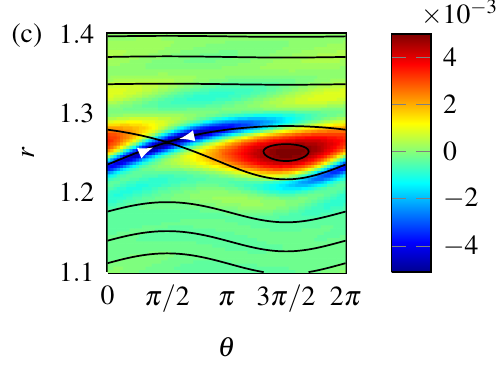}
%\tikzset{external/remake next}
%\input{newfigs3/1mode_phi_comp_1D.pgf}
%\hspace{2em}
%\tikzset{external/remake next}
%\input{newfigs3/1mode_phi_i_amp2.pgf}
%\tikzset{external/remake next}
%\input{newfigs3/1mode_phi_diff_amp2.pgf}
}
\caption{(a) Ideal and resistive forms of $\Delta \phi$ as in Fig.~\ref{fig:1mode_phi}a,b, but along the line $\theta=\pi/2$. Note the large peak in $\Delta \phi_i$ near the mode rational surface.
(b) $\Delta \phi_i(r,\theta)$ and (c) $\Delta \phi_i(r,\theta)-\Delta \phi_r(r,\theta)$, as in Figs.~\ref{fig:1mode_phi}a and \ref{fig:1mode_phi}c, but with mode amplitude $a=0.01$. Notice how the structure becomes concentrated near the stable manifold of the X-line.
}
\label{fig:1mode_phi_cont}
\end{center}
\end{figure}

We have computed $\Delta\tilde{\phi}_i$ by integrating Eq.~(\ref{eq:BdotGradPhi}) along field lines of the field ${\bf B}_0 + a_1\tilde{\bf B}$ from $z=-L$ to $z=L$, using $a=1\times 10^{-3}$ (see Fig.~\ref{fig:1mode_phi}).  This quantity is concentrated around the mode rational surface at $r_s=1.2576$, near both the X-line ($\theta\approx\pi/2$), and the elliptic (O) line ($\theta\approx 3\pi/2$). On both the hyperbolic and the elliptic closed field lines, $|\Delta \tilde{ \phi}_i|$ is large because the perturbation is constant along these lines.  For larger amplitude $a=0.01$, the island opens up, and $|\Delta \tilde{ \phi}_i|$ becomes localized around the stable manifold of the X-line (see Figs.~\ref{fig:1mode_phi_cont}b,c).
We have also shown the surfaces of the helical flux $\chi(r,\theta+kz)=\chi_0(r)+\tilde{\chi}(r,\theta+kz)=mA_z-krA_{\theta}$ at $z=-L$. Surface of section points for periodicity $L_p=2\pi/k$ lie on surfaces of constant $\chi$. 

The potential $\Delta \tilde{ \phi}_i$ can also be computed analytically. For infinitesimal $\tilde{\phi}_i$ and $\tilde{\mathbf{A}}$ we use $d\theta/dz=1/\mu(r)$ for the equilibrium fields to solve $\mathbf{B}_0\cdot\nabla\tilde{\phi}_i=-\gamma{\bf B}_0\cdot\tilde{\mathbf{A}}$.  With the normalization chosen so that $\tilde{A}_\theta(r)$ and $\tilde{A}_r(r)$ are pure imaginary, the potential difference is\footnote{The $r$ dependence of $\tilde{\bf A}$, ${\bf B}_{0}$, and $\mu$ is suppressed here for clarity.}:
\begin{eqnarray}
\Delta\tilde{\phi}_i(r,\theta_0)&=&\frac{\gamma{\mathbf B}_0\cdot \tilde{{\mathbf A}}}{B_{z0}}\int_{-L}^L dz ~\sin(m\theta(z)+kz)
\label{eq:sinccalc1}\\
&=&-\frac{\gamma{\mathbf B}_0\cdot \tilde{{\mathbf A}}}{B_{z0}} \left[\frac{\cos(m\theta_0+m(z+L)/\mu+kz)}{k+m/\mu}\right]_{-L}^L \\
&=& 2L\frac{\gamma{\mathbf B}_0\cdot \tilde{{\mathbf A}}}{B_{z0}}\sin(m(\theta_0+L/\mu))\frac{\sin\left((k+m/\mu)L\right)}{\left(k+m/\mu\right)L}\,\label{eq:sinccalc2}
\end{eqnarray}
where $\theta_0=\theta$ at $z=-L$. The quantity $\left(m/\mu(r)+k\right)={\bf k}\cdot {\bf B}_0 /B_{z0}$, so $\Delta\tilde\phi_i$ is proportional to a sinc function with a peak at the mode rational surface $r=r_s$. The width of $\Delta\tilde{\phi}_i$ in $r$ is 
\begin{equation}
w_m=\left|\frac{\pi \mu(r_s)^2}{mL\mu '(r_s)}\right|.\label{eq:SingleModeWidth}
\end{equation}
Note that the cosine factor in $\Delta \tilde\phi_i$ depends on both $\theta_0$ and $r$ [through $\mu(r)$], so its zero contours will be tilted in the $(r,\theta_0)$ plane.
For this calculation with infinitesimal mode amplitude, there is no information about the hyperbolic and elliptic lines, and in fact the island is assumed to have zero width. The width $w_m$ represents the decorrelation due to the shear in the magnetic field at the mode rational surface, which is proportional to $\mu '(r_s)$. This width is consistent with the observed radial width of $\Delta{\phi}_i$ in Fig.~\ref{fig:1mode_phi}a.  Note that in Fig.~\ref{fig:1mode_phi}a, the similar prefactor $({\bf B}\cdot \tilde{\bf A})/B_{z}$  goes rapidly to zero outside the mode rational surface, so the amplitude of $\Delta\phi_i$ falls off more rapidly than might be expected from the sinc function alone.  A plot of $\Delta\tilde\phi_i$ as given by Eq.~(\ref{eq:sinccalc2}) [including the prefactor $({{\mathbf B}_0\cdot \tilde{\mathbf A}})/{B_{z0}}$] is nearly indistinguishable from the plot in Fig.~\ref{fig:1mode_phi}a, which was computed by integrating Eq.~(\ref{eq:BdotGradPhi}a) numerically along field lines of ${\bf B}_0 + a\tilde{\bf B}$.

\begin{figure}[htbp]
\begin{center}
\commentfig{
\includegraphics[]{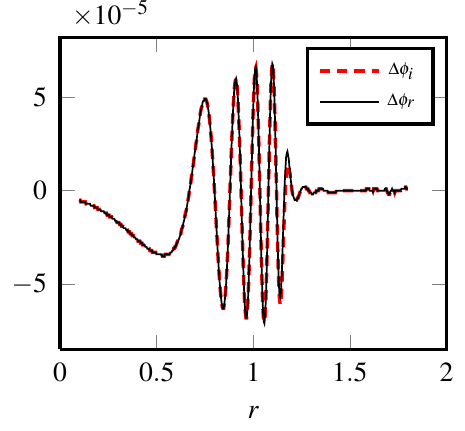}
%\tikzset{external/remake next}
%\input{newfigs3/ver2_1mode_phi.pgf}
}
\caption{Ideal and resistive forms of $\Delta \phi$ as in Fig.~\ref{fig:1mode_phi_cont}a, but with parameters $\nu$, $\eta$, $k_1$, and $L$  as in Case II (see Table \ref{tab:params}). While the difference between $\Delta \phi_i$ and $\Delta \phi_r$ is largest near the mode rational surface, it is still small compared to $\Delta \phi_i$.  This is an indication that mode width $w_m$ (Eq.~\ref{eq:SingleModeWidth}) is larger than the tearing width $w_t$ for this mode, and thus we expect that the effects of the line tying will be important.}
\label{fig:case2_1mode_phi}
\end{center}
\end{figure}

Fig.~\ref{fig:1mode_phi}b shows $\Delta \phi_r$, computed from Eq.~(\ref{eq:BdotGradPhi}b). The oscillations below $r=1.2$ are the same as in Fig.~\ref{fig:1mode_phi}a, but the larger peaks near $r=r_s$ have disappeared, because $-\gamma {\bf B} \cdot \tilde{\bf A}$ balances $\eta \tilde{\bf j} \cdot {\bf B}$ near $r=r_s$. The difference $\Delta\phi_i - \Delta\phi_r$ is plotted in Fig.~\ref{fig:1mode_phi}c, showing only the voltage drop due to the resistive term near $r=r_s$.
In Fig.~\ref{fig:1mode_phi_cont}a we show $\Delta \phi_i$ and $\Delta \phi_r$ as a function of $r$ for $\theta=\pi/2$. The large peak in near $r=r_s$ is evident in $\Delta \phi_i$ but not in $\Delta \phi_r$. We show $\Delta \phi_i$ in Fig.~\ref{fig:1mode_phi_cont}b, as in Fig.~\ref{fig:1mode_phi}a but with mode amplitude $a=0.01$.  In Fig.~\ref{fig:1mode_phi_cont}c we show the difference $\Delta\phi_i - \Delta\phi_r$ as in Fig.~\ref{fig:1mode_phi}c.  We see that the minimum and maximum values of $\Delta \phi_i$, at the X-line and the O-line, respectively, are equal and opposite, but the values near the X-line are more concentrated along the stable manifold of the X-line.  In Fig.~\ref{fig:case2_1mode_phi}, we show a plot as in Fig.~\ref{fig:1mode_phi_cont}a for $L=252$.  There is very little difference between $\Delta \phi_i$ and $\Delta \phi_r$ for this case.  As we will discuss in the next section, this is consistent with $w_m$ being larger than the tearing mode width $w_t$.

\section{Line-tied tearing modes}\label{sec:line-tied}

For analyzing modes with line-tying, we use the method of Ref.~\cite{evstatiev:072902}. We expand the perturbed velocity and magnetic field components $\xi$ as sums over the spectrum:
\begin{equation}\label{eq:spectrum}
\xi(r,\theta,z,t) = e^{im\theta + \gamma t}
\sum_k a_k \xi_k (r) e^{ikz},
\end{equation}
These \emph{radial eigenfunctions} satisfy the linearized equations (\ref{eq:mhdlin1}) and (\ref{eq:mhdlin2}) and the wavenumbers $k$ and the expansion coefficients $a_k$ are chosen so that the solution satisfies the line-tied boundary conditions at $z=\pm L$. If $L\gg 1$, then keeping only two terms $k_1$, $k_2$ in the sum gives a good approximation to the line-tied solution. This is the \emph{two mode approximation}. It is valid for large $L$ because $k_2-k_1$ is small for large $L$ and therefore $\xi_{k_1}(r) \simeq \xi_{k_2}(r)$. See Ref.~\cite{evstatiev:072902}. The calculation proceeds as follows.

The line-tied boundary conditions at $z=\pm L$ correspond to having the tangential components of $\tilde{{\bf E}}$ and the normal component of $\tilde{{\bf B}}$ be zero at $z=\pm L$. (We describe these modes as line-tied, in spite of the fact that the tangential velocity at $z=\pm L$  is nonzero due to finite resistivity.)   We proceed by treating $\gamma$ as fixed for now, and choose $k_1$ and $k_2$ such that in the infinite cylinder results $\gamma(k_1)=\gamma(k_2)$. Then the line-tying condition leads to $a_1 e^{ik_1 z}+a_2 e^{ik_2 z}=0$ at $z=\pm L$.
This implies that
$k_2(\gamma)-k_1(\gamma)=l \pi /L$, giving a dispersion relation for $\gamma$. Here $l$ is an integer, but we can take $l=1$ without loss of generality, and for $l=1$ we have $a_1=a_2\equiv a$. The length of our line-tied system is then $2L$, where $L = \pi/(k_2-k_1)$.

For this line-tied mode satisfying the two-mode approximation, we calculate the geometric width $w_g$ in the following manner\cite{delzanno:032904}. We find the two mode rational surfaces (${\bf k}\cdot {\bf B}_0=0$) corresponding to $k_1$ and $k_2$ by $m/\mu(r_1)+k_1=m/\mu(r_2)+k_2=0$.  The geometric width is the distance between these mode rational surfaces:
\begin{equation}
w_g\equiv r_2-r_1\simeq\left|\frac{(k_2-k_1)\mu(r_1)^2}{m\mu'(r_1)}\right|.\label{eq:geomwidth}
\end{equation}
In Ref.~\cite{delzanno:032904} it was shown that this definition is still qualitatively correct even for smaller lengths $L$, for which the two-mode approximation is not valid.  It is interesting to note that the single mode width $w_m$ in Eq.~(\ref{eq:SingleModeWidth})  equals $w_g$ if the line-tying relation $k_2-k_1=\pi/L$ is substituted.   We shall return to this point in the next section.

For $a$  large enough, the islands corresponding to the two modes will overlap and magnetic field line chaos will occur.\footnote{As in the last section, we will allow superposition of states with relatively large $a$ even when nonlinear effects are important. For sure, having field line chaos is such a nonlinear effect.} In order to be able to construct a surface of section plot with fields having a common period in $z$, we do the following: We first assume that $k_2/k_1= n_2/n_1$ is a rational number.  In this case the two modes then have a common length $L_p$ over which they are periodic,
$
L_p = n_1 L_1 = n_2 L_2,
$
where $L_i = 2\pi/|k_i|$.
Also, it can be seen from the line-tying relation above that 
$
L = \frac{1}{2}L_p/|n_2-n_1|.
$
So, if we take $|n_2-n_1|=1$, then $L_p = 2L$.  From this we conclude that the fields are actually periodic over the length of the line-tied system $z=-L \to L$. For this special case, the map from $z=-L$ to $z=L$ is a surface of section, and we can plot its return map to assess chaos in the periodic system.

Given the perturbed fields $\tilde{\bf B}$ from this two-mode calculation, we can find $\tilde{\bf A}$ in the gauge where $\tilde{A}_r =0$.  For the $k_1$ component, we obtain
\begin{eqnarray}
\tilde A_{z1}(r) &=& -\int^r_0 \tilde B_{\theta1}(r') \,dr',\\
\tilde A_{\theta1}(r) &=& \frac{m}{rk_1}\tilde A_{z1}(r) + \frac{i}{k_1} \tilde B_{r1}(r),\label{eq:Ath}
\end{eqnarray}
and similarly for the $k_2$ component.

The perturbed vector potential for the two mode approximation is thus
\begin{eqnarray}
\tilde{\bf A}(r,\theta,z,t) =a e^{\gamma t+im\theta}\left(\tilde{\bf A}_1(r)e^{ik_1 z} + \tilde{\bf A}_2(r)e^{ik_2 z}\right).
\end{eqnarray}
The inductive electric field ${\tilde{\bf E}}_I = -\partial_t {{\bf A}}$ is then:
\begin{eqnarray}
{\tilde{\bf E}}_I = -\gamma a e^{\gamma t+im\theta}\left(\tilde{{\bf A}}_1(r)e^{ik_1 z} + \tilde{{\bf A}}_2(r)e^{ik_2 z}\right).
\end{eqnarray}
This is the inductive field used for computing the scalar potentials $\Delta\phi_i$ and $\Delta\phi_r$ [c.f.~Eq.~(\ref{eq:BdotGradPhi})].  This mode behaves in $z$ as $e^{ik_0z} \cos(\delta k z)$, where $k_0$ is $(k_1+k_2)/2$ and $\delta k$ is $k_2-k_0$.  Therefore at $r=r_0\approx(r_1+r_2)/2$ the mode is nearly constant along the field lines for most of the length of the plasma.

\section{Reconnection diagnostics for line tied modes}

Given the two-mode approximation for line-tied modes, we use the perturbed fields $\tilde{\bf B}$ to compute the squashing factor $Q$ and the scalar potentials $\phi_i$ and $\phi_r$ of Eq.~(\ref{eq:BdotGradPhi}).  Two cases are compared: first, the plasma parameters are chosen so that the two-mode approximation is fairly accurate, and the tearing width $w_t$ (estimated from the width of the perturbed current near the mode rational surface) is larger than the geometric width $w_g$.  The second case has these limits reversed, with $w_t<w_g$; in this case, the two-mode approximation is not as accurate, but can be used as a simple model for understanding qualitative aspects of the line-tied mode.

\begin{table}[htdp]
\caption{Parameters for the two line-tied cases\label{tab:twomode} }
\begin{center}
%\begin{tabular}{c|c|c|c|c|c|c|c|c|c|c|c}
\begin{tabular}{rccccccccccc}
&$\eta$ & $\nu$ & $n_2/n_1$ & $k_1$ & $k_2$ & $\gamma$ 
& $w_g$ & $w_t$ & $L_1$ & $L_2$ & $L$ \\
\hline\hline
Case I: & $10^{-4}$ &$10^{-3}$& 19/20 & -0.135 & -0.128 & 0.0233 &0.047 &0.2&46.5&49.0&465 \\
Case II: & $10^{-6}$ &$10^{-5}$& 11/12 & -0.149 &-0.137 & 0.0110 & 0.084&0.05& 42.1&45.9& 252 \\
%Case I & $10^{-4}$ &$10^{-3}$& 19/20 & -0.1351 & -0.1283 & 0.0233 &0.047 &0.2&46.52&48.97&465.17 \\
%Case II & $10^{-6}$ &$10^{-5}$& 11/12 & -0.1494 &-0.1370 & 0.0110 & 0.084&0.05& 42.05&45.88& 251.87 \\
\end{tabular}
\end{center}
\label{tab:params}
\end{table}%

\subsection{Case I: $w_t>w_g$}

\begin{figure}[htbp]
\begin{center}
\commentfig{
\includegraphics[]{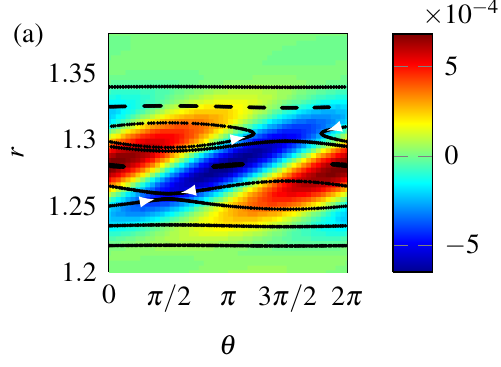}
\includegraphics[]{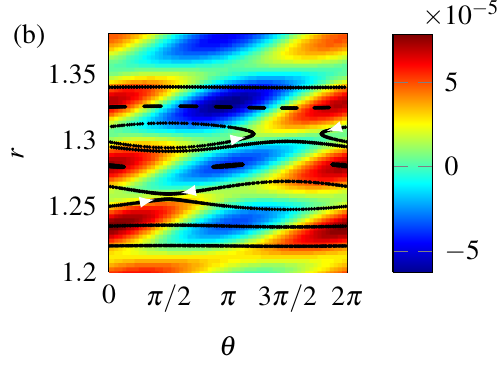}
\includegraphics[]{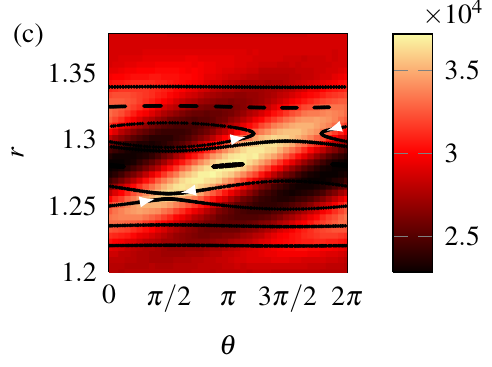}
%\tikzset{external/remake next}
%\input{newfigs3/2mode_phi_i.pgf}
%\tikzset{external/remake next}
%\input{newfigs3/2mode_phi_r.pgf}
%\tikzset{external/remake next}
%\input{newfigs3/2mode_Q.pgf}
}
\caption{Reconnection diagnostics for the two mode approximation with mode amplitude $a=1\times 10^{-3}$. All quantities computed with field lines mapped from $z=-L \to L$.
(a) $\Delta \phi_i(r,\theta)$, (b) $\Delta \phi_r(r,\theta)$, and (c) squashing factor $Q(r,\theta)$. Black dots show the surface of section map, which shows two sets of islands, and a secondary resonance.  Stable manifolds of the X-lines are indicated by the white arrows.
\label{fig:caseI.1}
}
\end{center}
\end{figure}

\begin{figure}[htbp]
\begin{center}
\commentfig{
\includegraphics[]{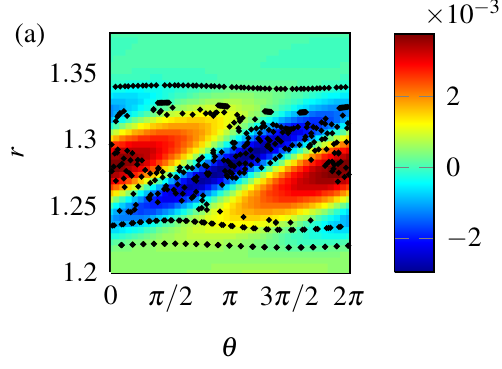}
\includegraphics[]{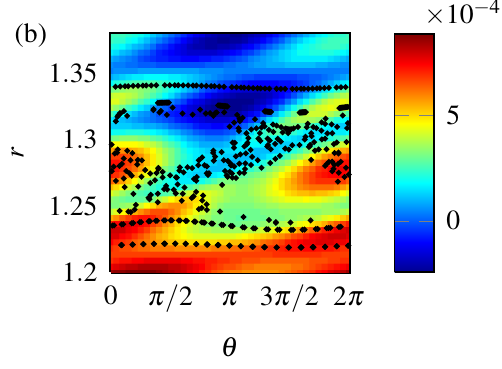}
\includegraphics[]{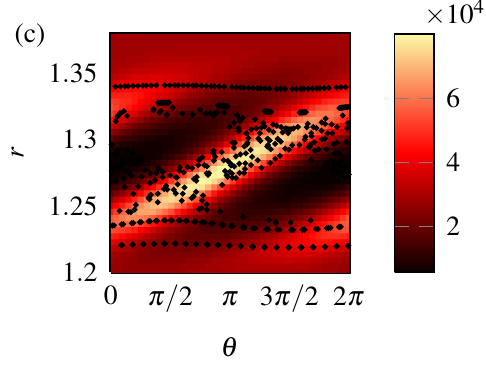}
%\tikzset{external/remake next}
%\input{newfigs3/2mode_phi_i_amp2.pgf}
%\tikzset{external/remake next}
%\input{newfigs3/2mode_phi_r_amp2.pgf}
%\tikzset{external/remake next}
%\input{newfigs3/2mode_Q_amp2.pgf}
}
\caption{Reconnection diagnostics for the two mode approximation with mode amplitude $a=5\times 10^{-3}$. All quantities computed with field lines mapped from $z=-L \to L$.
(a) $\Delta \phi_i(r,\theta)$, (b) $\Delta \phi_r(r,\theta)$, and (c) squashing factor $Q(r,\theta)$. Black dots show the surface of section map, showing field line chaos due to the overlap of the two sets of islands.  The two X-lines are still near $\theta =\pi /2,r=r_1$ and $\theta=3 \pi /2, r=r_2$, respectively.
\label{fig:caseI.2}
}
\end{center}
\end{figure}

The plasma parameters for this case are $\eta = 1\times10^{-4}$ and $\nu = 1\times10^{-3}$. The two modes have $k_2/k_1 = 19/20$, $L=465$, and the values of $k_i$, $\gamma$,
and $L_i$ are given in Table \ref{tab:twomode}. Two different mode amplitudes are shown for this case.  In Fig.~\ref{fig:caseI.1}, the mode amplitude is $a=1\times10^{-3}$.  The surface of section shows two sets of islands with very thin secondary resonances between them.  The $\Delta\phi_i$ calculation (Fig.~\ref{fig:caseI.1}a) shows structure near the islands for the two modes. The fact noted in the last section, just after Eq.~(\ref{eq:geomwidth}), namely that $w_m = w_g$, implies that the regions of large $|\Delta \phi_i|$ from the two island chains overlap. This is true because $w_g$, as defined in Eq.~(\ref{eq:geomwidth}), is the radial displacement between the island chains. The values of $|\Delta \phi_i|$ are peaked near the secondary resonances with $r=r_0$, between $r=r_1$ and $r=r_2$, and aligned with the stable manifolds of the X-lines of the two island chains.  On the other hand, $\Delta\phi_r$ (Fig.~\ref{fig:caseI.1}b) is much smaller in magnitude and not localized. The squashing factor is also localized near the islands (Fig.~\ref{fig:caseI.1}c).  Figure \ref{fig:caseI.2} shows results for the same two modes, but with amplitude $a=5\times10^{-3}$.  Note that the surface of section now shows a chaotic region linking the islands, and $\Delta\phi_i$ (Fig.~\ref{fig:caseI.2}a) and squashing factor $Q$ (Fig.~\ref{fig:caseI.2}c) are both now more localized in the chaotic region and aligned with the stable manifolds of the X-lines. The resistive potential $\Delta \phi_r$ is smaller in magnitude and shows little variation in the island region (Fig.~\ref{fig:caseI.2}b).

\subsection{Case II: $w_t<w_g$}

\begin{figure}[htbp]
\begin{center}
\commentfig{
\includegraphics[]{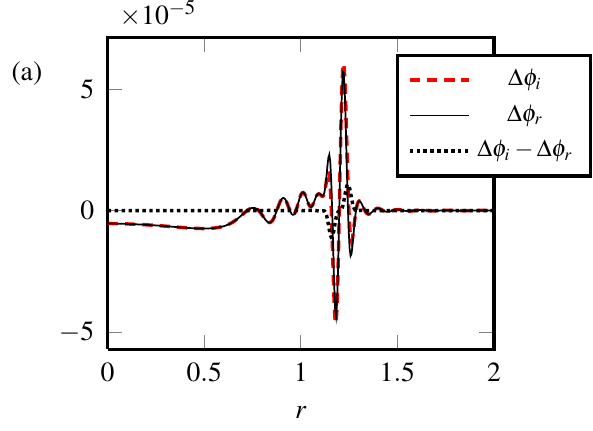}
\hspace{2em}
\includegraphics[]{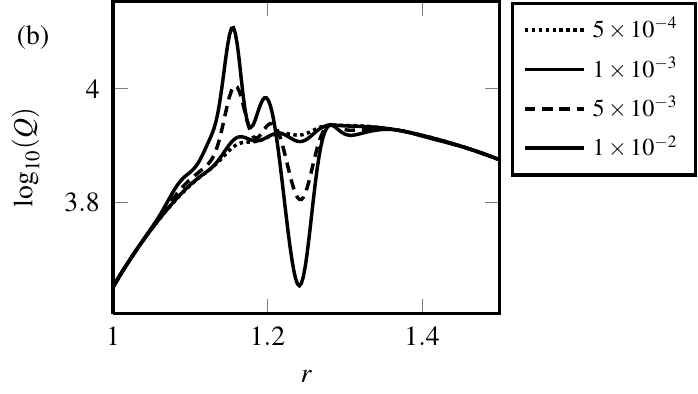}
%\tikzset{external/remake next}
%\input{newfigs3/ver2_2mode_phi.pgf}
%\hspace{2em}
%\tikzset{external/remake next}
%\input{newfigs3/ver2_2mode_phi_diff.pgf}
%\hspace*{2em}
%\tikzset{external/remake next}
%\input{newfigs3/ver2_2mode_Q.pgf}
}
\caption{
Reconnection diagnostics for the case $w_t<w_g$. (a) $\Delta\phi_i$, $\Delta\phi_r$, and a measure of the parallel current integrated along field lines, given by $\Delta\phi_i-\Delta\phi_r$ ($a=10^{-3}$, $\theta=\pi/2$).  In this case, the finite length of the system is more important than the parallel current, as can be seen in (a), since $\Delta\phi_i$ and $\Delta\phi_r$ are nearly the same.  
In (b) is shown ${\rm log}_{10}(Q)$ for various mode amplitudes $a$. The peaks in $Q$ are quite evident for large $a$ but not noticeable for small $a$.
The importance of the finite length ($w_t < w_g$) is not obvious from $Q$.
\label{fig:caseII}
}
\end{center}
\end{figure}

The plasma parameters for this case are $\eta = 1\times10^{-6}$ and $\nu = 1\times10^{-5}$. The two modes have $k_2/k_1 = 11/12$, $L=252$, and the values of $k_i$, $\gamma$, and $L_i$ are given in Table \ref{tab:twomode}.
For this shorter system, $\Delta \phi_i$ and $\Delta \phi_r$ are nearly equal throughout the plasma (Fig.~\ref{fig:caseII}a), meaning that the resistive voltage drop, proportional to $\Delta \phi_i - \Delta \phi_r$, is very small. This is an indication that the mode has lost its tearing character because of the shorter length of the plasma.
This effect is quantified in Ref.~\cite{delzanno:032904}, where it was shown that the scaling of the growth rate is tearing-like (with $\gamma\propto$ a fractional power of $\eta$) when $w_t>w_g$ as in the previous case, but the mode exhibits resistive diffusion (with $\gamma \propto \eta$) whenever $w_t < w_g$.
Even in the region where the current is strong (the tearing layer) the finite length is a more important factor than the influence of the current.
On the other hand, the squashing factor (Fig.~\ref{fig:caseII}b) is quite localized in the region of the two island chains, indicating the stretching character of the X-lines of the islands. In this case, the variation of $Q$ in this region suggests the potential for reconnection, whereas the potentials $\Delta \phi_i$ and $\Delta \phi_r$ show that for this case there is essentially no reconnection going on. In contrast with $Q$, the slip-squashing factor of Ref.~\cite{0004-637X-693-1-1029} measures the actual field line slippage due to non-ideal effects as well as stretching.

\section{Summary and conclusions}

We have constructed a cylindrical model for fields with a growing
mode in the presence of line-tying at $z=\pm L$. The fields in this
model consist of an equilibrium plus perturbed
fields $\tilde{\mathbf{E}}$, $\tilde{\mathbf{B}}$ of an arbitrary
but small amplitude. The perturbed fields are computed using the two-mode approximation. The two infinite
cylinder modes, with distinct axial mode numbers $k_{1}$, $k_{2}$,
are both tearing modes, so the line-tied mode formed from them should
have tearing character\cite{delzanno:032904,huang:042102} (i.e.~effectively
involve reconnection) for large $L$. Also, it is in this limit that
the two-mode approximation is valid.  Refs.~\cite{delzanno:032904,huang:042102} showed that for shorter plasmas, the modes no longer have tearing character.  We have traced field lines and
computed the squashing factor $Q$ and the electrostatic potentials
$\Delta \phi_i$, $\Delta \phi_r$ obtained from Eq.~(\ref{eq:BdotGradPhi}), where the inductive
electric field is given in terms of the linear modes by $\tilde{\mathbf{E}}_I=-\gamma\tilde{\mathbf{A}}$
and $\gamma$ is the linear growth rate.
This model, with an equilibrium plus linear modes of arbitrary amplitude,
is accurate for very small mode amplitudes but clearly inaccurate
for large amplitudes, for which the nonlinear terms are important.
Quantitative results for large amplitude nonlinear states require
nonlinear resistive MHD simulations, which are outside the scope of the present investigation.

We discussed the point that the squashing factor or squashing ratio depends on only the geometry of the magnetic fields, whereas the electrostatic potentials $\Delta \phi_i$ and $\Delta \phi_r$ involve the magnetic field geometry plus the inductive electric fields. So, large and localized values of $Q$ indicate a magnetic structure with the potential for reconnection; large and localized values of $\Delta\phi_i$ relative to $\Delta\phi_r$ indicate electric and magnetic fields for which reconnection should be occurring. A comparison between our methods using $\Delta\phi_i$ and $\Delta\phi_r$  and the use of the slip-squashing factor of Ref.~\cite{0004-637X-693-1-1029} should be interesting, but is also outside the scope of the present study.

We found very different behavior for $\Delta \phi_i$, $\Delta \phi_r$ in Case I, the long plasma case ($w_g<w_t$) and Case II, the short plasma case ($w_g>w_t$). In Case I, $\Delta\phi_i$ is much more peaked in the tearing layer region than $\Delta \phi_r$, indicative that reconnection plays a large role in the line-tied mode in this range. In Case II, $\Delta \phi_i$ and $\Delta \phi_r$ are basically equal, indicating that the shorter length precludes reconnection. These results are consistent with the results of \cite{delzanno:032904,huang:042102}, where $w_g$ and $w_t$ have the same respective meanings as in this paper. In that earlier work, it was found that the growth rates $\gamma$ scale as the appropriate fractional power of $\eta$ for Case I and scale linearly with $\eta$ for Case II. 
Both $Q$
and $\Delta\phi_i$ showed localized but nonsingular structures in the areas
in which the corresponding periodic system has hyperbolic closed field
lines or (for larger amplitude modes) chaotic behavior. On the other hand, the squashing factor shows similar behavior for Cases I and II. In particular, it shows almost no variation near the mode islands for small amplitude modes and localized behavior for larger mode amplitudes, with no clear signature of the different mode structure in Case I and Case II.

\section*{Acknowledgments}

We wish to thank V.~Titov, E.~Zweibel, and V.~Mirnov for valuable suggestions. This research was supported by the DOE Office of Science,
Fusion Energy Sciences and performed under the auspices of
the NNSA of the U.S. DOE by LANL, operated by LANS
LLC under Contract No. DE-AC52-06NA25396.

\section*{Appendix: Issues related to the squashing factor $Q$}

As discussed in Sec.~\ref{sec:diagnostics}, the squashing factor is found in the following
manner. First, let $J$ be the Jacobian matrix for the map $\mathbf{x}\rightarrow\mathbf{X}$ which takes field lines from one surface, where $B_n<0$, to the other surface, where $B_n>0$. These surfaces are ${z=-L}$ and ${z=L}$, respectively, in our example.
For the special case $B_{z0}={\rm const.}$, we have
$D=\det(J)=1$. The squashing factor is then simply the trace of $J^{T}J$, as given by Eq.~(\ref{eq:squashing}) with
$D=1$.  For the more general flux preserving case
with $B_{z}$ not constant, the determinant is $D=B_{z}(x_{1},x_{2},-L) /B_{z}(X_{1},X_{2},L) \neq 1 $. It is tempting to change to canonical variables, i.e.~variables ${\bf u}$ and ${\bf U}$ in which the equations of motion are canonical and the map is area
preserving. This is always possible by the Darboux theorem, and it may appear at first blush that the situation is simplified because the determinant in these variables is unity.

However, for the physical problems we consider, there is a natural metric on the two surfaces in the original variables, obtained from the Euclidean metric in 3D, so lengths are written as $(\delta{\bf x},G \delta{\bf x})$. (For our example, the surfaces $z=-L$ and $z=L$ are flat, so this metric is the identity, $G=I$.) For general $G$, the relevant Rayleigh quotient is

\begin{equation}\label{eq:RayQuo-1}
Q_R\equiv\frac{(\delta{\bf x},J^T G J \delta{\bf x})}{(\delta{\bf x},G \delta{\bf x})},
\end{equation}
where $J_{ij}={\partial X_{i}}/{\partial x_{j}}$. The relevant eigenvalues are the eigenvalues of $G^{-1}J^T G J$.

The metric in the new canonical variables inherits the metric $(\delta{\bf u},\tilde{G} \delta{\bf u})=(\delta{\bf x},G \delta{\bf x})$. This implies that $\tilde{G}$ satisfies $M^T \tilde{G} M=G$, where $M$ is the tangent map from $\delta {\bf x}$ to $\delta{\bf u}$ (and from $\delta {\bf X}$ to $\delta{\bf U}$). The new Rayleigh quotient is
\begin{equation}\label{eq:RayQuo-2}
\tilde{Q}_R\equiv\frac{(\delta{\bf u},\tilde{J}^T \tilde{G} \tilde{J} \delta{\bf u})}{(\delta{\bf u},\tilde{G} \delta{\bf u})},
\end{equation}
which equals $Q_R$. Therefore, if we transform to canonical variables and inherit the metric from the (physically motivated) metric on the original surface, we do not gain any simplicity from the relation $D=1$.

This issue is related to the issue that arises for finite time Lyapunov exponents (FTLE), where the interval $-L<z<L$ is replaced by the time interval $0<t<T$, and ${\bf x}\rightarrow {\bf X}(t=T)$. If we change variables but continue to use the Euclidean metric, the FTLE 
\begin{equation}
h(T)=\frac{1}{T}\rm{ln} \left[ \frac {(\delta \mathbf{X}(T),\delta \mathbf{X}(T)) }{(\delta \mathbf{x},\delta \mathbf{x})}\right]
\end{equation}
are not coordinate invariant. The infinite time (largest) Lyapunov exponent, defined as $h=\lim_{T\rightarrow \infty} h(T)$,
\emph{is} coordinate invariant. The more natural question to ask is whether there is a natural metric. If so, then the metric changes upon a change of variables ${\bf x}\rightarrow {\bf u}$ and ${\bf X}\rightarrow {\bf U}$ so the FTLE are invariant as above. If there is no natural metric, then the FTLE depend on the metric, i.e.~the Rayleigh quotient $(\delta \mathbf{x},J^T G J \delta \mathbf{x}) / (\delta \mathbf{x},G\delta \mathbf{x})$ depends on the metric $G$. The infinite time Lyapunov exponents, however, are independent of $G$.

A final issue relates to the angle between the magnetic field and the two surfaces. If ${\bf B}$ is far from normal to the surface, a circular flux tube projects to an elongated ellipse on the surface, and this is not related to reconnection. This issue was dealt with in Ref.~\cite{0004-637X-660-1-863}. For the model used in this paper, $B_{z0} \gg B_{\theta0}$ and the magnetic field is almost normal to the surfaces $z=-L$ and $z=L$, so that this issue is not important.

\bibliographystyle{unsrt}
%\bibliography{refs}

\begin{thebibliography}{10}

\bibitem{delzanno:032904}
G.~L. Delzanno and J.~M. Finn.
\newblock The effect of line-tying on tearing modes.
\newblock {\em Physics of Plasmas}, 15(3):032904, 2008.

\bibitem{huang:042102}
Y.-M. Huang and E.~G. Zweibel.
\newblock Effects of line tying on resistive tearing instability in slab
 geometry.
\newblock {\em Physics of Plasmas}, 16(4):042102, 2009.

\bibitem{Priest:1995fk}
E.~R. Priest and P.~D{\'e}moulin.
\newblock Three-dimensional magnetic reconnection without null points, 1. basic
 theory of magnetic flipping.
\newblock {\em J. Geophys. Res.}, 100(A12):23443--23463, 1995.

\bibitem{1996A&A...308..643D}
P.~{D{\'e}moulin}, J.~C. {Henoux}, E.~R. {Priest}, and C.~H. {Mandrini}.
\newblock {Quasi-Separatrix layers in solar flares. I. Method.}
\newblock {\em Astron. Astrophys.}, 308:643--655, April 1996.

%\bibitem{1999A&A...351..707T}
%V.~S. {Titov} and P.~{D{\'e}moulin}.
%\newblock {Basic topology of twisted magnetic configurations in solar flares}.
%\newblock {\em A\&A}, 351:707--720, November 1999.

\bibitem{Titov20021087}
V.~S. Titov and G.~Hornig.
\newblock Magnetic connectivity of coronal fields: geometrical versus
 topological description.
\newblock {\em Advances in Space Research}, 29(7):1087 -- 1092, 2002.

\bibitem{1990ApJ...350..672L}
{Y.-T.} {Lau} and J.~M. {Finn}.
\newblock {Three-dimensional kinematic reconnection in the presence of field
 nulls and closed field lines}.
\newblock {\em The Astrophysical Journal}, 350:672--691, February 1990.

\bibitem{1991ApJ...366..577L}
{Y.-T.} {Lau} and J.~M. {Finn}.
\newblock {Three-dimensional kinematic reconnection of plasmoids}.
\newblock {\em The Astrophysical Journal}, 366:577--591, January 1991.

\bibitem{1988JGR....93.8583G}
J.~M. {Greene}.
\newblock {Geometrical properties of three-dimensional reconnecting magnetic
 fields with nulls}.
\newblock {\em Journal of Geophysical Research}, 93:8583--8590, August 1988.

\bibitem{NR:BirnPriestBook}
J.~Birn and E.~Priest, eds.
\newblock {\em Reconnection of Magnetic Fields; Magnetohydrodynamics
and Collisionless Theory and Observations}, chapter
 2, pages 16-86
\newblock Cambridge University Press, 2007.

\bibitem{PhysRevLett.96.015004}
W.~F. Bergerson, C.~B. Forest, G.~Fiksel, D.~A. Hannum, R.~Kendrick, J.~S.
 Sarff, and S.~Stambler.
\newblock Onset and saturation of the kink instability in a current-carrying
 line-tied plasma.
\newblock {\em Phys. Rev. Lett.}, 96(1):015004, Jan 2006.

\bibitem{furno:2324}
I.~Furno, T.~Intrator, E.~Torbert, C.~Carey, M.~D. Cash, J.~K. Campbell, W.~J.
 Fienup, C.~A. Werley, G.~A. Wurden, and G.~Fiksel.
\newblock Reconnection scaling experiment: A new device for three-dimensional
 magnetic reconnection studies.
\newblock {\em Review of Scientific Instruments}, 74(4):2324--2331, 2003.

\bibitem{4989217}
W.~Gekelman, H.~Pfister, Z.~Lucky, J.~Bamber, D.~Leneman, and J.~Maggs.
\newblock Design, construction, and properties of the large plasma research
 device-the {LAPD} at {UCLA}.
\newblock {\em Review of Scientific Instruments}, 62(12):2875 --2883, December
 1991.

\bibitem{Hesse:1988rt}
M.~Hesse and K.~Schindler.
\newblock A theoretical foundation of general magnetic reconnection.
\newblock {\em J. Geophys. Res.}, 93(A6):5559--5567, 1988.

\bibitem{Schindler:1988yq}
K.~Schindler, M.~Hesse, and J.~Birn.
\newblock General magnetic reconnection, parallel electric fields, and
 helicity.
\newblock {\em J. Geophys. Res.}, 93(A6):5547--5557, 1988.

\bibitem{PhysRevLett.103.105002}
E.~E. Lawrence and W.~Gekelman.
\newblock Identification of a quasiseparatrix layer in a reconnecting
 laboratory magnetoplasma.
\newblock {\em Phys. Rev. Lett.}, 103(10):105002, Sep 2009.

\bibitem{ESASpec.Publ.448:7151999}
V.~S. {Titov}, P.~D{\'e}moulin, and G.~Hornig.
\newblock Quasi-separatrix layers: Refined theory and its application to solar
 flares.
\newblock In et~al. A.~Wilson, editor, {\em ESA SP-448, Magnetic Fields and
 Solar Processes}, page 715, 1999.

\bibitem{Titov:2002fj}
V.~S. Titov, G.~Hornig, and P.~D{\'e}moulin.
\newblock Theory of magnetic connectivity in the solar corona.
\newblock {\em J. Geophys. Res.}, 107(A8), 08 2002.

\bibitem{0004-637X-660-1-863}
V.~S. Titov.
\newblock Generalized squashing factors for covariant description of magnetic
 connectivity in the solar corona.
\newblock {\em The Astrophysical Journal}, 660(1):863, 2007.

\bibitem{0004-637X-693-1-1029}
V.~S. Titov, T.~G. Forbes, E.~R. Priest, Z.~Miki{\'c}, and J.~A. Linker.
\newblock Slip-squashing factors as a measure of three-dimensional magnetic
 reconnection.
\newblock {\em The Astrophysical Journal}, 693(1):1029, 2009.

\bibitem{evstatiev:072902}
E.~G. Evstatiev, G.~L. Delzanno, and J.~M. Finn.
\newblock A new method for analyzing line-tied kink modes in cylindrical
 geometry.
\newblock {\em Physics of Plasmas}, 13(7):072902, 2006.

\bibitem{NR:svd_evals}
W.~H. Press, S.~A. Teukolsky, W.~T. Vetterling, and B.~P. Flannery.
\newblock {\em Numerical Recipes: The Art of Scientific Computing}, chapter
 11.0.6, pages 569--570.
\newblock Cambridge University Press, third edition, 2007.

\bibitem{APD:currentsandHFT}
{G. Aulanier}, {E. Pariat}, and {P. D\'emoulin}.
\newblock Current sheet formation in quasi-separatrix layers
 and~hyperbolic~flux~tubes.
\newblock {\em A\&A}, 444(3):961--976, 2005.

\bibitem{0004-637X-582-2-1172}
V.~S. Titov, K.~Galsgaard, and T.~Neukirch.
\newblock Magnetic pinching of hyperbolic flux tubes. i. basic estimations.
\newblock {\em The Astrophysical Journal}, 582(2):1172, 2003.

\bibitem{0004-637X-631-2-1227}
M.~Hesse, T.~G. Forbes, and J.~Birn.
\newblock On the relation between reconnected magnetic flux and parallel
 electric fields in the solar corona.
\newblock {\em The Astrophysical Journal}, 631(2):1227, 2005.

\bibitem{richardson:112511}
A.~S. Richardson, J.~M. Finn, and G.~L. Delzanno.
\newblock Control of ideal and resistive magnetohydrodynamic modes in reversed
 field pinches with a resistive wall.
\newblock {\em Physics of Plasmas}, 17(11):112511, 2010.

\end{thebibliography}
%\mycomment{%%

%}%% end comment

\end{document}